\documentclass[fleqn,usenatbib]{mnras}

\usepackage{newtxtext,newtxmath}
\usepackage{amsmath}
\usepackage{graphics,graphicx}
\usepackage{natbib}
\usepackage{caption,setspace} 

\usepackage{bm}

\usepackage{mathtools}
\usepackage{enumitem}
\usepackage{lipsum}
\setlist[itemize]{leftmargin=*}

\usepackage{floatrow}
\usepackage[outercaption]{sidecap}    
\sidecaptionvpos{figure}{t}

\usepackage[T1]{fontenc}
\usepackage{ae,aecompl}
\pdfoutput=1
\usepackage{xspace}
\newcommand{\skirt}{{\sc skirt}\xspace}
\newcommand{\statmorph}[0]{\texttt{statmorph} }
\newcommand{\fsps}[0]{\texttt{fsps} }

\title[The sizes and stellar mass surface densities of massive FIRE-2 galaxies]{Realistic mock observations of the sizes and stellar mass surface densities of massive galaxies in FIRE-2 zoom-in simulations}

\author[T. Parsotan et al.]{T. Parsotan,$^{1}$\thanks{E-mail: parsotat@oregonstate.edu}
R. K. Cochrane,$^{2}$
C. C. Hayward, $^{3}$
D. Angl\'es-Alc\'azar, $^{3,4}$
R. Feldmann,$^{5}$\newauthor
C. A. Faucher-Gigu\`ere, $^{6}$ 
S. Wellons, $^{6}$
and P. F. Hopkins $^{7}$ \\
\vspace{0.1cm}\\
$^{1}$Department of Physics, Oregon State University, 301 Weniger Hall, Corvallis, OR 97331, USA\\
$^{2}$Harvard - Smithsonian Center for Astrophysics, 60 Garden St, Cambridge, MA 02138, USA\\
$^{3}$Center for Computational Astrophysics, Flatiron Institute, 162 Fifth Avenue, New York, NY 10010, USA  \\
$^{4}$Department of Physics, University of Connecticut, 196 Auditorium Road, U-3046, Storrs, CT 06269-3046, USA \\
$^{5}$Institute for Computational Science, University of Zurich, Zurich CH-8057, Switzerland\\
$^{6}$Department of Physics \& Astronomy and CIERA, Northwestern University, 1800 Sherman Ave, Evanston, IL 60201, USA\\
$^{7}$TAPIR, MC 350-17, California Institute of Technology, Pasadena, CA 91125, USA}
\date{Accepted XXX. Received YYY; in original form ZZZ}

\pubyear{2020}
\begin{document}
\label{firstpage}
\pagerange{\pageref{firstpage}--\pageref{lastpage}}
\maketitle

\begin{abstract} 
The galaxy size-stellar mass and central surface density-stellar mass relationships are fundamental observational constraints on galaxy formation models. However, inferring the physical size of a galaxy from observed stellar emission is non-trivial due to various observational effects, such as the mass-to-light ratio variations that can be caused by non-uniform stellar ages, metallicities, and dust attenuation. Consequently, forward-modeling light-based sizes from simulations is desirable. In this work, we use the {\skirt} dust radiative transfer code to generate synthetic observations of massive galaxies ($M_{*}\sim10^{11}\,\rm{M_{\odot}}$ at $z=2$, hosted by haloes of mass $M_{\rm{halo}}\sim10^{12.5}\,\rm{M_{\odot}}$)
from high-resolution cosmological zoom-in simulations that form part of the Feedback In Realistic Environments (FIRE) project. The simulations used in this paper include explicit stellar feedback but no active galactic nucleus (AGN) feedback. From each mock observation, we infer the effective radius ($R_e$), as well as the stellar mass surface density within this radius and within $1\,\rm{kpc}$ ($\Sigma_e$ and $\Sigma_1$, respectively). We first investigate how well the intrinsic half-mass radius and stellar mass surface density can be inferred from observables. The majority of predicted sizes and surface densities are within a factor of two of the intrinsic values. We then compare our predictions to the observed size-mass relationship and the $\Sigma_1-M_\star$ and $\Sigma_e-M_\star$ relationships. At $z\gtrsim2$, the simulated massive galaxies are in general agreement with observational scaling relations. At $z\lesssim2$, they evolve to become too compact but still star-forming, in the stellar mass and redshift regime where many of them should be quenched. Our results suggest that some additional source of feedback, such as AGN driven outflows, is necessary in order to decrease the central densities of the simulated massive galaxies to bring them into agreement with observations at $z\lesssim2$. \newline
\end{abstract}

\section{Introduction}
Observations of distant galaxies are crucial for understanding the physics orchestrating galaxy evolution and the assembly of galaxy structures (see \citealt{Conselice2014}, for a review). The period around the peak of cosmic star formation, around $1\lesssim z \lesssim3$, is particularly important; at this epoch, stellar mass is building most rapidly (see the review by \citealt{Madau2014} and references therein), and measuring galaxy structure here can provide constraints on the drivers of high star formation rates. In particular, structures and morphologies can help distinguish between models of star formation (`inside-out' versus `outside in' growth; e.g. \citealt{vanDokkum2010,Wuyts2012,Dokkum2015,Tacchella2016,Tacchella2018,Whitney2019,Spilker2019}), determine the relative importance of in-situ star formation as opposed to merger-driven mass assembly \citep{Stott2011,Newman2012,Huertas2015,Hill2017,Hill2019} and discriminate between quenching mechanisms \citep{Wu2018,Wu2020,Wang2019}. However, characterisation of the structures of high redshift galaxies has historically been challenging, due to the small angular sizes of distant galaxies and the resolution limitations of ground-based telescopes.\\
\indent Space-based imaging, notably the Hubble Space Telescope ({\it{HST}}), has been critical to the development of this field. Deep data, in particular from the Cosmic Assembly Near-infrared
Deep Extragalactic Legacy Survey (CANDELS; \citealt{Grogin2011,Koekemoer2011}) has the necessary combination of high angular resolution (of order $0.1-0.2''$) and sensitivity to infer typical sizes of massive galaxies to $z\sim7$ \citep{Allen2017,Hill2017}. At low and intermediate redshifts ($0 \lesssim z\lesssim3$), more detailed analysis has been possible, and lower stellar mass galaxies can be studied. It is now well-established that galaxy size correlates with properties such as stellar mass, star formation rate and color, and that empirical scaling relations evolve with redshift. More massive galaxies are, on average, larger than less massive ones, both in the local Universe \citep{Shen2003,Lange2015} and at high redshift \citep{Trujillo2004,Barden2005}. At fixed stellar mass and redshift, star-forming galaxies are larger than their quiescent counterparts, at least out to $z\sim2$ (e.g. \citealt{Toft2009,Williams2010,Barro2017,Whitaker2017}). At high redshift, galaxies tend to be more compact \citep{Ferguson2004,Daddi2005,Buitrago2008}, with the most significant size evolution observed for galaxies classified as quiescent (e.g. \citealt{Williams2010,Carollo2013,Mosleh2017}). These various correlations are encapsulated in the evolving size-mass relations (e.g. \citealt{van_der_Wel_CANDELS_size_mass}, though see \citealt{suess2019half} for extensive discussion of the pitfalls of observational measurements of galaxy size).\\
\indent Stellar surface density (e.g. within the innermost $1\,\rm{kpc}$) is also observed to be correlated with various galaxy properties.  Massive, quiescent galaxies tend to have higher stellar surface densities, with less dense galaxies displaying higher star-formation rates, on average \citep{Franx2008,Williams2010,Whitaker2017}. These relations also evolve with redshift; at fixed stellar surface density, galaxies at higher redshifts are more highly star-forming \citep{Franx2008}. \\
\indent As observations have provided a clearer view of the history of stellar mass assembly, simulations have attempted to explain observational results and use them to constrain their sub-grid models for key physical processes such as feedback from stars and massive black holes. One important question that has been explored is how AGN feedback leaves an imprint on the physical sizes of galaxies and on their central densities \citep{Fan2008,Dubois2013,Ishibashi2013,Wellons2015,Genel2018,Vlugt2019}. \cite{choi2018} recently explored this with two sets of simulations, one with and one without AGN feedback (though including stellar feedback). They showed that the galaxies simulated with AGN feedback showed a suppression of central cooling, resulting in lower stellar mass density in their cores. Similarly, \cite{Appleby2019} show that the X-ray black hole feedback implemented in the SIMBA cosmological hydrodynamical simulations \citep{Dave2019} pushes dense gas outwards, lowering the central specific star formation rate. \cite{Zoldan2019} also argue that quasar-driven mechanical winds are required to reconcile simulations with observed galaxy sizes. Therefore, AGN feedback appears to be required not only to quench star formation in massive galaxies (e.g. \citealt{Somerville2015}), but also to regulate their sizes and central densities.  However, most current cosmological simulations rely on extensive tuning of sub-grid parameters to match observations, which limits their predictive power.\\
\indent Another key limitation of using simulations to interpret observational results lies in the lack of observable predictions made by most simulations. For example, while studies such as \cite{choi2018} compared the sizes of their galaxies to observationally-derived relations between stellar mass and surface density, they typically do not fully forward-model their simulations for direct comparison with observations. Cosmological simulations do not, in general, fold the details of dust geometry into their output, and providing predictions for simulated galaxies with all possible observational setups (given the numerous variables, such as telescope, waveband, seeing, and instrument noise) would be impossible. However, interest in this field is growing, with accessible radiative transfer software \citep[e.g.][]{Jonsson2006,Jonsson2010,Dullemond2012,SKIRT} enabling mock observables to be generated with relative ease \citep[e.g.][]{Hayward2014,HS2015,Trayford2017,Camps2018,cochrane2019predictions,Liang2018,Liang2019,Ma2019}.\\
\indent In this paper, we evaluate the extent to which stellar feedback alone can regulate the sizes and central densities of the most massive galaxies in the Feedback In Realistic Environments 2 (FIRE-2) cosmological `zoom-in' simulations \citep{Hopkins2017} \footnote{\url{http://fire.northwestern.edu}} presented in \cite{angles2017BH_fire}.  FIRE simulations include a variety of stellar feedback physics implemented explicitly in a multi-phase interstellar medium (ISM), and have been shown to reproduce the size-mass relation at $z=0$ for $M_{*}<10^{10.5}\,\rm{M_{\odot}}$ \citep{El_Badry_2016}, the Kennicutt-Schmidt relation \citep{Orr2018}, and the mass-metallicity relation \citep{Ma2016}. In this work, we probe the limits of stellar feedback in the extreme environments of the inner kpc of massive galaxies ($M_{*}\sim10^{11}\,\rm{M_{\odot}}$) at $z=1-3$.\\
\indent We build on the work performed by \cite{price2017testing}, who test how well the sizes and stellar masses of FIRE galaxies can be recovered using mock images. {We note a few key differences between their work and ours here. Firstly, while \cite{price2017testing} made use of the MassiveFIRE suite of galaxies \citep{Feldmann2016, Feldmann_2017}, simulated using the original FIRE module, we use updated FIRE-2 physics and a novel implementation of supermassive black hole (SMBH) accretion and growth, but neglect AGN feedback entirely. In this paper, we put particular emphasis on our projection of the simulations into `observer-space', including the effects of dust attenuation. \cite{price2017testing} applied a \cite{calzetti2000dust} dust attenuation curve to individual stellar particles, so that the effective attenuation depended on the line-of-sight density of dust (or metals). We implement a more sophisticated model for dust attenuation and re-emission, via three-dimensional continuum radiative transfer, and also model projection effects. This enables us to simulate multi-wavelength emission in a self-consistent manner, accounting for the geometry of the dust and star particles. Like \cite{price2017testing},} we generate broadband images and convolve these with typical telescope point spread function. We then analyse the resultant mock observations in the same way as real data. This involves fitting each mock observation with a S\'ersic profile, and deriving the effective radius, the mass-to-light ratio, and the stellar mass surface density. {\cite{price2017testing} tested the recovery of intrinsic FIRE galaxy sizes at $z\sim2$. Here, we extend these tests to a wider range of redshifts ($1.25<z<2.76$), and additionally test the recovery of the stellar mass surface density. Further extending the previous study, we make direct comparisons to the observationally-derived scaling relations presented by \cite{van_der_Wel_CANDELS_size_mass} and \cite{Barro2017}}.\\
\indent The structure of the paper is as follows. In Section \ref{sec:sims}, we describe the FIRE-2 simulations and outline the creation of mock observations. In Section \ref{sec:methods}, we describe the methods used to measure stellar mass surface densities and effective radii and present the results of the analysis (with additional plots presented in the Appendix). In Section \ref{sec:discussion}, we discuss the implications of our findings. We present our conclusions in Section \ref{sec:conclusions}.

\section{A sample of simulated high-redshift galaxies}\label{sec:sims}
\subsection{Four massive, central galaxies from the FIRE-2 simulations}\label{sec:fire_sims}
\noindent The FIRE project \citep{Hopkins2014,Hopkins2017} is a set of state-of-the-art hydrodynamical cosmological zoom-in simulations. One of the key motivations for these simulations was a more complete understanding of the role of stellar feedback in galaxy evolution. Stellar feedback is believed to regulate star-formation and the masses of galaxies over time. In particular, it is needed to match observationally-inferred gas consumption timescales (e.g. \citealt{Hopkins2011}), galaxy stellar mass functions \citep[e.g.][]{Davidzon2017} and the stellar mass-to-halo mass relation \citep{Moster2010,Moster2013,Behroozi2013,Cochrane2017a}, as well as to explain the metal enrichment of the circumgalactic medium and intergalactic medium (e.g. \citealt{Oppenheimer2006,Muratov2017,Hafen2019}).\\
\indent The FIRE project reaches sufficient mass and force resolution to model various stellar feedback processes including supernovae, photo-heating, stellar mass loss from O- and AGB-stars and radiation pressure \citep[see][]{Dale2015} directly. The simulations do this explicitly via two main methods. The first is resolving the formation of giant molecular clouds (GMCs). Star formation in the FIRE simulations takes place in self-gravitating (according to the \citealt{Hopkins2013sf_criteria} criterion), self-shielding molecular gas (see \citealt{Krumholz2011}) at high densities ($n_{H}>1000\,\rm{cm}^{-3}$ in the simulations used in this paper). The second involves modelling mass, metal, energy, and momentum return using the predictions of stellar population synthesis (SPS) models, without explicit parameter tuning, which is necessarily applied in large-volume cosmological simulations. The details of the feedback mechanisms implemented are presented in \cite{Hopkins2018feedback}. The simulations have been broadly successful at generating galactic winds self-consistently \citep{Angles_Alcazar_2017_baryon_cycle, Muratov2017} and reproducing observed galaxy properties, such as stellar masses, star-formation histories, metallicities, morphologies and kinematics \citep{Hopkins2014,VandeVoort2015,Ma2016,Feldmann_2017,Ma2017,Sparre2017}.\\
\indent In this paper, we focus on the four central galaxies of the massive haloes simulated by \citep{angles2017BH_fire} using the FIRE-2 model \citep{Hopkins2017}. The haloes were first simulated by \citet{Feldmann2016, Feldmann_2017} using the original FIRE model \citep{Hopkins2014}, as part of the {\sc MassiveFIRE} suite. {Compared to FIRE, our new FIRE-2 simulations are run with a more accurate hydrodynamics solver (a mesh-free Godunov solver implemented in the {\sc gizmo}
\footnote{\url{http://www.tapir.caltech.edu/~phopkins/Site/GIZMO.html}} 
code; \citealt{Gaburov2011,Hopkins_2015_GIZMO}). They also feature improved treatments of cooling and recombination rates, gravitational softening and numerical feedback coupling, and they adopt a higher density threshold for star-formation \citep{Hopkins2018feedback}. Our simulations include a new treatment for the seeding and growth of SMBHs via gravitational torque-driven accretion (though no AGN feedback); see \citet{Angles-Alcazar2013,Angles-Alcazar2015,angles2017BH_fire} for details.} The mass resolution is $3.3\times{}10^4$ $M_\odot$ for gas and star particles and $1.7\times{}10^5$ $M_\odot$ for dark matter particles. We denote our simulated central galaxies using their halo names, A1, A2, A4, and A8. At $z=2$, these haloes have masses of $M_{\rm{halo}}\sim10^{12.5}\,\rm{M_{\odot}}$ and host central galaxies with stellar masses of $7\times10^{10}-3\times10^{11}M_{\odot}$ and a range of assembly histories. A detailed kinematic analysis of these galaxies was presented in \cite{Wellons2019}.

\subsection{Post-processing with {\small SKIRT}}
The FIRE-2 simulations do not make direct predictions for observed emission. In order to make mock images of these galaxies, we must model the intrinsic stellar emission, and then the propagation of that emission between the source and the observer. To do this, we use the radiative transfer methods \citep{Steinacker2013} implemented in the Stellar Kinematics Including Radiative Transfer ({\skirt})\footnote{\url{http://www.skirt.ugent.be}} Monte Carlo radiative transfer code \citep{Baes2011,SKIRT}. Our methods are detailed in \cite{cochrane2019predictions}, where we presented a detailed analysis of the spatially-resolved dust continuum emission in the central galaxies of halos A1, A2, A4, and A8. We provide a brief description of the procedure here.\\
\indent We assign spectral energy distributions (SEDs) to the stars in each galaxy according to their ages and metallicities, using {\sc starburst99} templates \citep{starburst99} (these templates are also used in the FIRE simulations themselves), using a \cite{Kroupa2001} initial mass function (IMF). We model dust within the galaxy using a dust-to-metals mass ratio of $0.4$ \citep{Dwek1998,James2002}, assuming that dust is destroyed in gas particles with temperature $>10^6\,{\rm K}$ \citep{Draine1979,Tielens1994}. We use a \cite{WeingartnerDraineDust} Milky Way dust prescription, which includes a mixture of graphite, silicate and PAH grains. {\skirt} then tracks the paths of photons through this model dust distribution, tracking dust absorption (and self-absorption), scattering, and re-emission. \\
\indent We place detectors at five different angles with respect to the face-on galaxy to create mock observations at various inclinations. The inclinations are: $0^\circ, 60^\circ, 90^\circ, 120^\circ,$ and $180^\circ$, where $0^\circ$ is a face-on view of the halo, defined with respect to the angular momentum vector of the galaxy's gas particles, and $90^\circ$ is an edge-on view of the halo. This allows us to incorporate observational uncertainties that may arise due to viewing angle effects into our analysis.\\
\indent We perform this post-processing on a subset of snapshots spanning the peak of cosmic star-formation, when stellar mass is building up very rapidly \citep{Madau2014}. The redshifts studied are $z=1.25$, $z=1.75$, $z=2.25$, and $z=2.76$.

\section{Mock observations}\label{sec:methods}
\begin{figure*}[]
\includegraphics[width=\columnwidth]{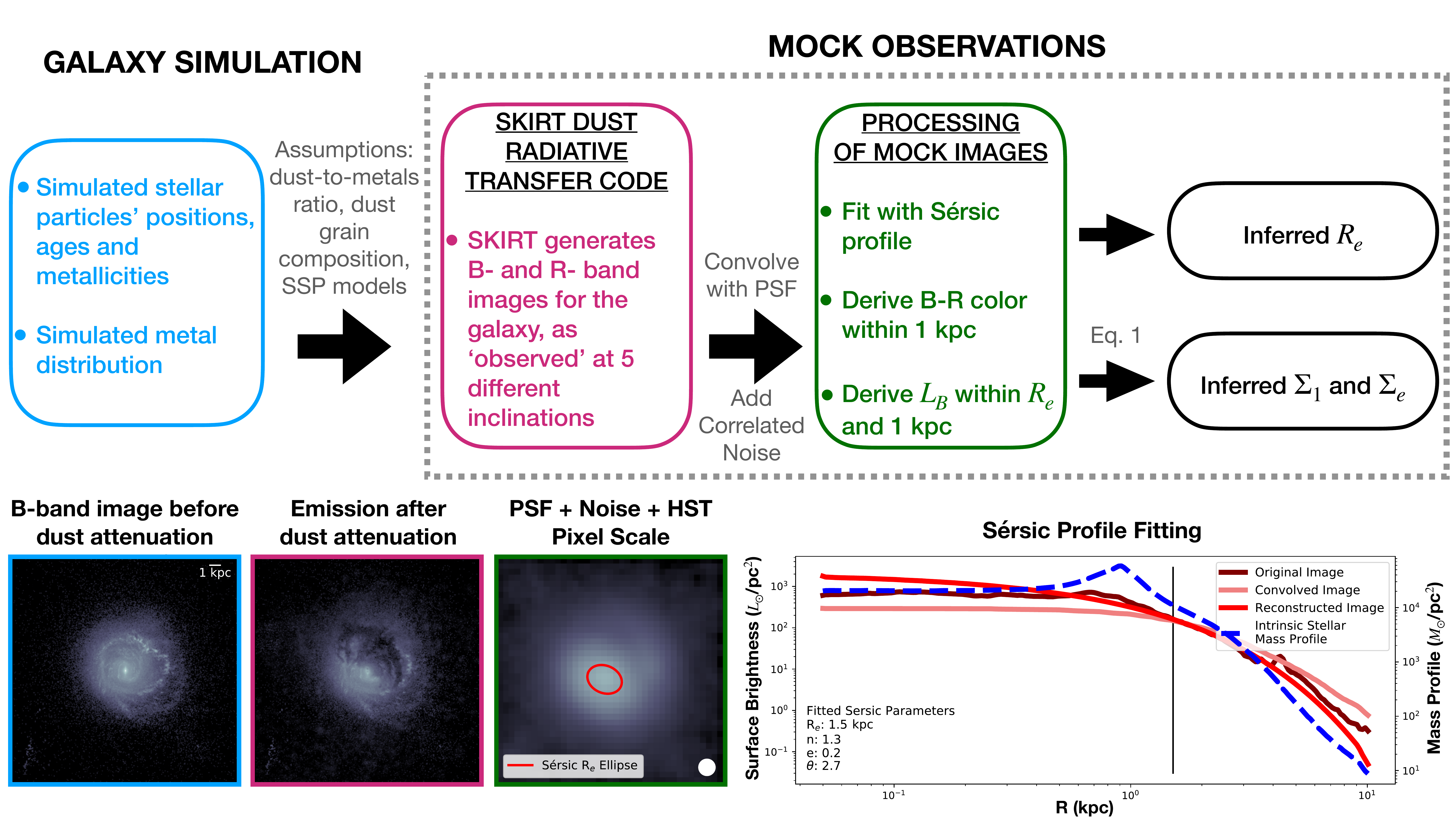}
\caption{The upper panel shows the workflow of this paper, and the lower panel shows an example of the process for an individual galaxy snapshot (galaxy A2, at $z=1.75$, with an face-on orientation). The three images show the transformation of a $B$-band galaxy image from dust-free image (left), to dust-attenuated image produced using radiative transfer (center), to the convolved image projected onto the HST 0.06'' drizzled pixel scale (right; beam size shown in white). The right-hand image includes background correlated noise that would be observed in the rest-frame $B$-band at the HST WFC/IR angular resolution. All three images have the same flux scale. The lower right-hand panel shows surface brightness profiles of the dust-unattenuated image (maroon) and the dust-attenuated, PSF-convolved image with noise (pink). The reconstructed surface brightness profile (red) is derived using S\'ersic profile fits to the convolved image. The S\'ersic profile fits are typically able to account for $\gtrsim90$\% of the light in the dust-attenuated, unconvolved image. The vertical black line shows the best-fitting stellar effective radius at this inclination, $1.5\,\rm{kpc}$. The dashed blue line shows the stellar mass surface density profile derived from a two-dimensional projection of the stellar mass particle data.}
\label{fig:example_fitting_flow}
\end{figure*}

\subsection{Deriving the sizes and surface densities of FIRE-2 galaxies}\label{sec:sizes_surface_densities}
We create mock photometric observations of each of the {\skirt} outputs, at each inclination, using the Johnson B and Kron-Cousins R filter transmission functions.\footnote{\url{http://svo2.cab.inta-csic.es/theory/fps/}} We also produce rest-frame $\sim 5000$\AA\ images, to match the rest-frame wavelength at which galaxy sizes are inferred in observational studies \citep[e.g.][]{van_der_Wel_CANDELS_size_mass}.
{We convolve the resultant images with the Hubble Space Telescope (HST) WFC3/IR point spread function (PSF) that was acquired by \cite{skelton2014_HST_PSF}.
For simplicity, we use this PSF for all images, following \cite{price2017testing}. 
We then resample each image to the HST WFC3 drizzled pixel scale of 0.06'' and insert it into a blank region of a random CANDELS HST F160W image \citep{skelton2014_HST_PSF,price2017testing}}.
This process yields mock observed images with realistic correlated noise (see Figure \ref{fig:example_fitting_flow}, for an example of the workflow).\\
\indent We then perform an analysis analogous to that used in observational studies. We fit a S\'ersic profile \citep{sersic} to each mock $B$-band image using the \statmorph python code \citep{Rodriguez-Gomez2019}. This procedure takes into account the PSF in order to fit for the intrinsic light distribution of the galaxy. We obtain the best-fitting ellipticity, angle of rotation, and the semi-major and semi-minor axes. The effective radii ($R_e$) quoted in this paper are the semi-major axes of the fitted Sersic ellipses, to match the definition used by the observational work we compare to \citep{van_der_Wel_CANDELS_size_mass}. Typically, the integrated S\'ersic profiles recover $\gtrsim$90\% of the light in the unconvolved {\skirt} images. Comparisons of the surface brightness profiles of unconvolved $B$-band image to the best-fit S\'ersic profile show that the fit is also able to reconstruct the surface brightness profile (Figure \ref{fig:example_fitting_flow}, bottom panel). In deriving the stellar effective radius in this way, we implicitly assume that there are no spatial variations in the mass-to-light ratio, consistent with the majority of observational analyses and in line with the results of \cite{price2017testing}. We discuss the limitations of this approach in Section \ref{sec:discussion}.\\
\indent We infer stellar mass surface densities from our synthetic images using well-established observational techniques. We follow the method outlined in \cite{bell2001_m_l_relation} to infer $B$-band mass-to-light ratios ($M/L_\mathrm{B}$) from observed $B-R$ color within $1\,\rm{kpc}$ and $R_e$ apertures. We derive $B-R$ colors using the methods established by \cite{Tacchella_2015_use_sersic_color}, which they show minimises the effects of the PSF on the result. We use $B$-band and $R$-band S\'ersic fits to calculate the flux at each of the two wavelengths, within a $1\,\rm{kpc}$ or $R_e$ aperture (note that we repeated this procedure using $R_{80}$, the radius which contains 80\% of the galaxy's light, and found very similar results. {{This reflects the flat mass-to-light ratios seen in these simulated galaxies}}). We then derive stellar mass within the aperture, using the $B$-band light (within an elliptical aperture with a semi-major axis of $1\,\rm{kpc}$ or the measured $R_e$), the calculated $B-R$ color and an updated \cite{bell2001_m_l_relation} relation (see Appendix \ref{sec:appendix_ml}). We repeat the process for galaxy images generated using different sky orientations to obtain an estimate of the standard deviation due to projection effects. This procedure enables us to calculate the stellar mass surface densities within $1\,\rm{kpc}$ and our measured $R_e$ ($\Sigma_1$ and $\Sigma_e$, respectively; see \citealt{Cheung2012}) \emph{that an observer would infer from the synthetic images}.\\
\indent {The total stellar mass is derived directly from the simulation particle data, using a sphere of radius $0.1R_{\rm{vir}}$, where $R_{\rm{vir}}$ is the virial radius of each galaxy. In principle, biases in recovering $M_\star$ would affect our comparisons with observations. However, a full investigation of this is beyond the scope of this paper, and various studies have found that $M_\star$ can be recovered within $\sim0.3\,\rm{dex}$ (see e.g. \citealt{HS2015,price2017testing,bagpipes}).}

\subsection{Recovery of intrinsic sizes and surface densities}\label{sec:size_recovery}
Before embarking on the main analysis of this paper, we study how well the inferred effective stellar radii reflect the intrinsic half-mass radii calculated directly from the massive galaxy simulations (in three dimensions, using spherical shells). We find that the stellar effective sizes measured from the synthetic galaxy images tend to be slightly larger than the half-mass sizes calculated directly from the simulation particle data. This is the case for 13 of our 16 snapshots (see Figure \ref{size_vs_size}). Nevertheless, the majority (12/16) of our inferred sizes are within a factor of two of the intrinsic size, defined as the half-mass radius derived from the 3-dimensional stellar particle data. {The median values of $\log_{10}(R_{e,\rm{inferred}}/R_{e,\rm{intrinsic}})$ are $0.15\,\rm{dex}$, $0.15\,\rm{dex}$, $0.32\,\rm{dex}$, and $-0.02\,\rm{dex}$, at $z=1.25$, $z=1.75$, $z=2.25$, and $z=2.75$, respectively. Across all haloes, redshifts and inclinations, the median offset is $0.17\,\rm{dex}$, with a standard deviation of $0.20\,\rm{dex}$.} The largest discrepancy between intrinsic and inferred galaxy size is seen at $z=2.25$. This is driven by galaxy A4, which at this redshift is clumpy and quite obscured by dust (see \citealt{cochrane2019predictions} for a more in-depth analysis of this amorphous morphology). This image is particularly difficult for \statmorph to fit. This is also an issue for galaxy A8 at $z=1.75$. \\
\indent Intrinsic stellar mass surface densities are also fairly well recovered from mock observations for the majority of snapshots. {The intrinsic stellar mass surface densities are acquired directly from the FIRE-2 simulations by calculating the total stellar mass within a given sphere, with a radius corresponding to $1\,\rm{kpc}$ or $R_{e,\rm{intrinsic}}$, and then dividing by projected surface area (e.g. $\Sigma_{e,\rm{intrinsic}}=M(R_{e,\rm{intrinsic}})/\pi R_{e,\rm{intrinsic}}^{2}$}). The median values of $\log_{10}(\Sigma_{1,\rm{inferred}}/\Sigma_{1,\rm{intrinsic}})$ are: $0.01\,\rm{dex}$, $0.07\,\rm{dex}$, $-0.20\,\rm{dex}$, and $0.18\,\rm{dex}$, at $z=1.25$, $z=1.75$, $z=2.25$, and $z=2.75$, respectively. Across all haloes, redshifts and inclinations, the median offset between $\Sigma_{1,\rm{inferred}}$  and $\Sigma_{1,\rm{intrinsic}}$ is $0.04\,\rm{dex}$, with a standard deviation of $0.37\,\rm{dex}$. The corresponding median values of $\log_{10}(\Sigma_{e,\rm{inferred}}/\Sigma_{e,\rm{intrinsic}})$ are: $-0.15\,\rm{dex}$, $-0.15\,\rm{dex}$, $-0.70\,\rm{dex}$, and $0.13\,\rm{dex}$. Across all haloes, redshifts and inclinations, the median offset between $\Sigma_{e,\rm{inferred}}$  and $\Sigma_{e,\rm{intrinsic}}$ is $-0.17\,\rm{dex}$, with a standard deviation of $0.52\,\rm{dex}$. The discrepancy between the intrinsic and inferred surface densities of galaxy A4 at $z=2.25$ (see Section \ref{sec:appendix_comp}, Figure \ref{density_vs_density_1kpc}) is due to the same effects that extend $R_e$ by a factor of $\sim 3$. \\
\indent We also consider the uncertainties due to inclination effects explicitly (these uncertainties correspond to the size of the error bars, $\sigma_{R_{e,\rm{inferred}}}$, $\sigma_{\Sigma_{e,\rm{inferred}}}$, and $\sigma_{\Sigma_{1,\rm{inferred}}}$, shown in Figures \ref{size_vs_size} and \ref{density_vs_density_1kpc}). We first calculate the percentage uncertainties on the inferred radii ($100\times \sigma_{R_{e,\rm{inferred}}}/R_{e,\rm{inferred}}$), and derive the mean percentage uncertainty of the four haloes at each redshift. These are $20\%$, $14\%$, $19\%$, and $24\%$, at $z=1.25$, $z=1.75$, $z=2.25$, and $z=2.75$. Next, we repeat the process for the inferred stellar mass surface densities. The percentage uncertainties on $\Sigma_{1}$ (i.e. mean of $100\times \sigma_{\Sigma_{1,\rm{inferred}}}/\Sigma_{1,\rm{inferred}}$) are $28\%$, $30\%$, $30\%$, and $43\%$, at $z=1.25$, $z=1.75$, $z=2.25$, and $z=2.75$. For $\Sigma_{e}$, the corresponding values are $37\%$, $29\%$, $31\%$, and $63\%$, at $z=1.25$, $z=1.75$, $z=2.25$, and $z=2.75$. As we will discuss in Section \ref{sec:obs_size_mass_results}, such inclination effects will increase the scatter in observed relations relative to intrinsic ones.

\begin{figure*}[]
\includegraphics[scale=1.0]{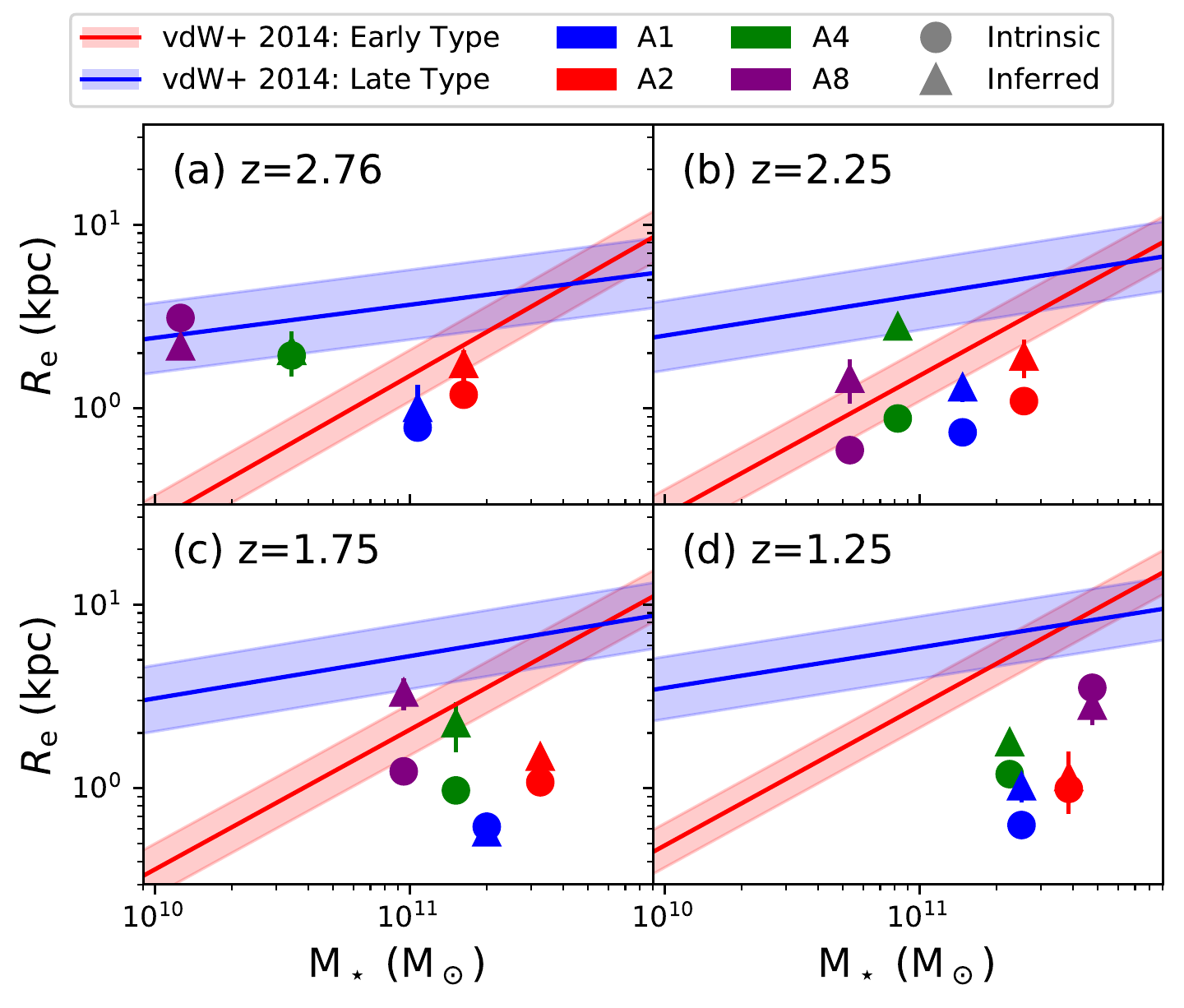} 
\caption{The stellar effective radius as a function of stellar mass, for each central galaxy at (a) $z=2.76$, (b) $z=2.25$, (c) $z=1.75$ and (d) $z=1.25$. We show the intrinsic half-mass radius (derived directly from the 3-dimensional distribution of stellar mass within the simulation, using spherical shells), as well as the effective radius derived from our mock observations (defined as the semi-major axis of the ellipse that contains half of the total flux of the integrated best-fitting S\'ersic model). Error bars are derived using the $1\sigma$ uncertainty on the measurements using simulated galaxies with five sky orientations. We overplot the $R_e-M_\star$ scaling relations \citep{van_der_Wel_CANDELS_size_mass} for both late-type (blue) and early-type (red) galaxies, with shaded regions showing the $1\sigma$ scatter. Values of $R_e$ obtained from fits to the $0.5\,\mu\rm{m}$ mock FIRE-2 observations fall below the late-type empirical relations for galaxies A1 and A2 at all redshifts. 
Our predictions for the observed sizes of galaxies A4 and A8 are in agreement with the late-type galaxy size-mass relationship at $z=2.75$, but these galaxies become too compact at lower redshifts.}
\label{vdW}
\end{figure*} 

\begin{figure*}[]
\includegraphics[scale=1.0]{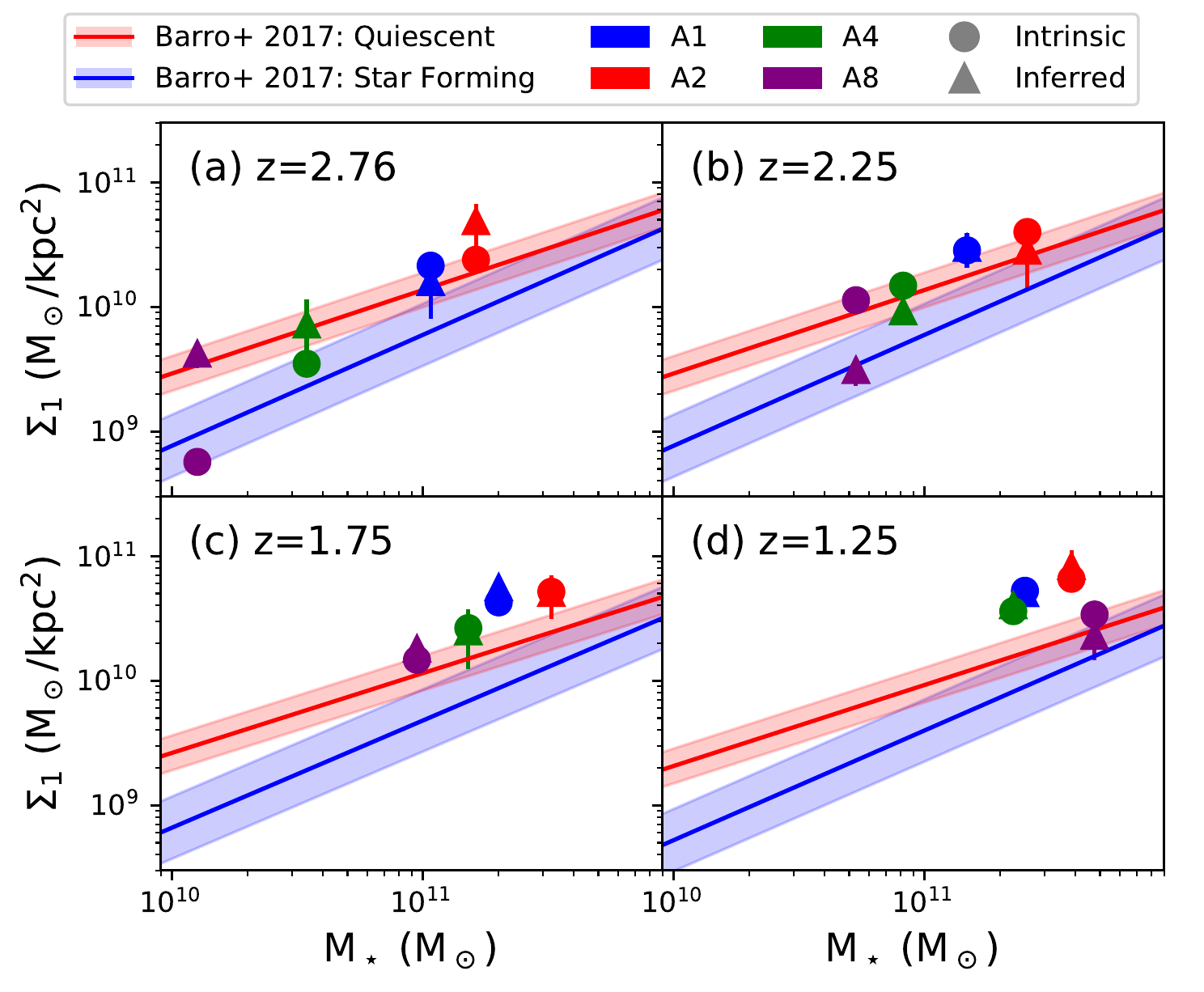} 
\caption{The stellar mass surface density, calculated within the central $1\rm{kpc}$, as a function of stellar mass, for each central galaxy at (a) $z=2.76$, (b) $z=2.25$, (c) $z=1.75$ and (d) $z=1.25$. As in Figure \ref{vdW}, the circles show the intrinsic values, derived directly from the simulation, and the triangles show the values inferred from our mock observations. Error bars are derived using the $1\sigma$ uncertainty on the measurements using simulated galaxies with five sky orientations. We overplot the empirical $\Sigma_1-M_\star$ scaling relations \citep{Barro2017}, with shaded regions showing the $1\sigma$ dispersion. At high redshifts, the FIRE-2 galaxies show general agreement with the scaling relation for star-forming galaxies. However, by $z=1.75$ the halos have begun to diverge from the star-forming scaling relation and by $z=1.25$, they lie a factor of $2$ above the empirical relation for quiescent galaxies, with the exception of galaxy A8.}
\label{barro_1}
\end{figure*}

\begin{figure*}[]
\includegraphics[scale=1.0]{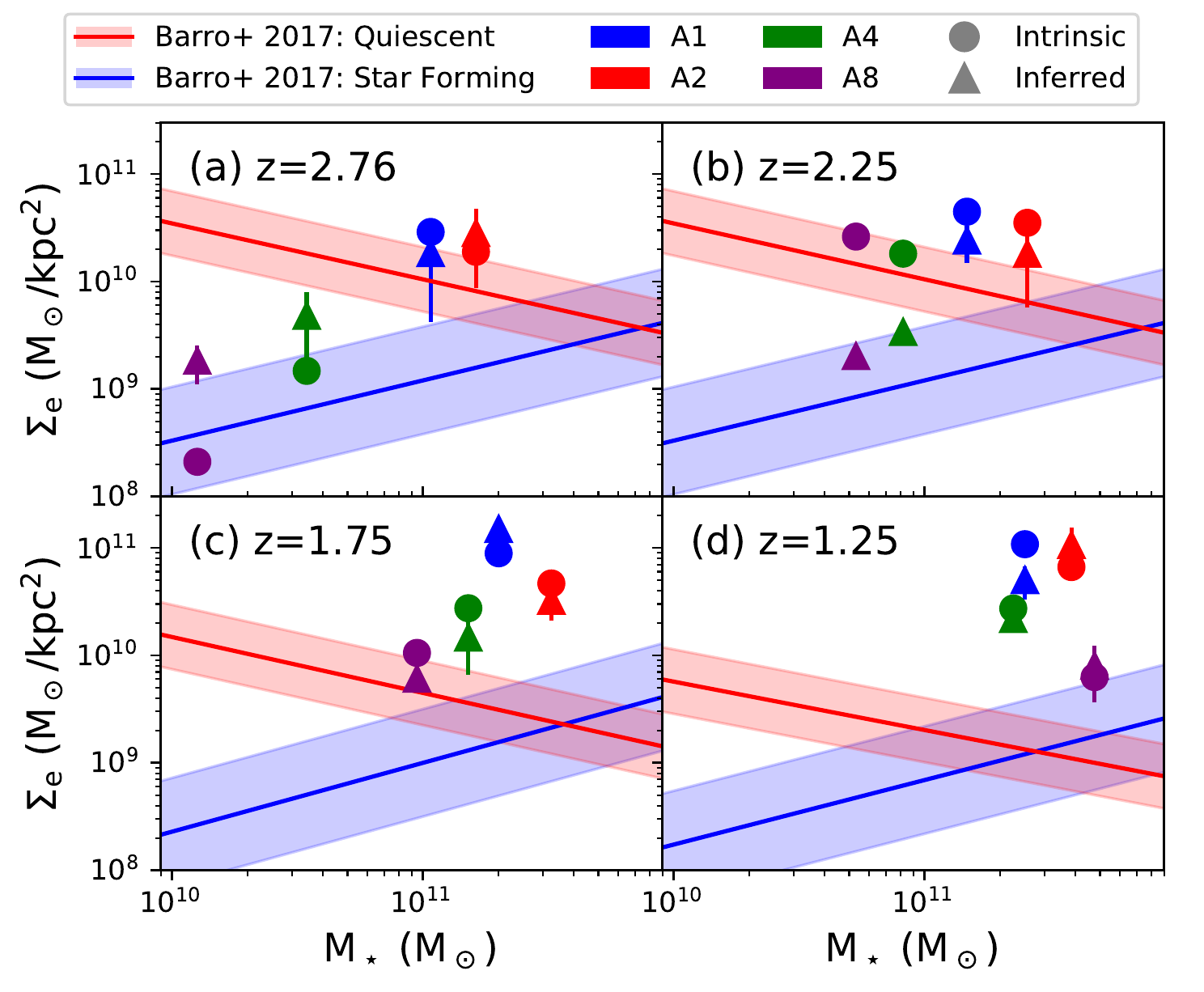} 
\caption{As Figure \ref{barro_1}, but with the stellar mass surface density calculated using the stellar effective radius $R_{e}$, rather than the central $1\,\rm{kpc}$ for the ``inferred'' values and the 3D calculated half-mass radii for the ``intrinsic'' values. Galaxies A4 and A8 are in agreement with the empirical scaling relation for star-forming galaxies at $z=2.76$, and galaxies A1 and A2 lie on the relation for quiescent galaxies. With the exception of galaxy A8, the galaxies diverge from these relations with time, lying over an order of magnitude above them by $z=1.25$.}
\label{barro_e}
\end{figure*}

\subsection{Comparison to observational size-mass relations}\label{sec:obs_size_mass_results}
In Figure \ref{vdW}, we show our measurements of the four massive FIRE-2 galaxies on the size-mass plane, at each of the four redshifts studied. We overplot the size-mass relation derived by \cite{van_der_Wel_CANDELS_size_mass}, who also use rest-frame $0.5\,\mu$m images. We find that the closest agreement between the massive FIRE-2 galaxies and the observationally-derived relation occurs at high redshifts. At $z=2.76$, two of the four halos are broadly consistent with the late-type galaxy size-mass relation, and two are broadly consistent with the early-type relation. Note that, based on UVJ rest-frame colors, these FIRE-2 galaxies would be classed as star-forming at all snapshots studied here. This is expected, since AGN feedback, which is believed to play a role in the quenching of galaxies, is not included in these simulations. At lower redshifts, the agreement worsens. By $z=1.25$, all of the simulated galaxies are significantly offset below the observationally-derived \cite{van_der_Wel_CANDELS_size_mass} relations for both early and late-type galaxies.\\
\indent One interesting feature of these results is the difference between intrinsic and inferred sizes. As noted in Section \ref{sec:size_recovery}, the inferred sizes are generally within a factor of two of those calculated directly from the simulation data. Yet the empirical relations are fairly tight, and, in some cases, the differences between intrinsic and observed sizes are larger than the scatter in the empirical relations. The morphology/viewing angle of the sources, as quantified by the error bars on each of the data points, contributes to this. The difference between intrinsic and inferred size could have implications for studies of the scatter in scaling relations, in particular for work that attempts to reproduce this scatter in simulations. Our results suggest that proper forward-modelling of simulations into observational space is necessary for the scatter in scaling relations of simulated galaxies to be interpreted in a meaningful way.

\subsection{Comparison to observational surface density-mass relations}\label{sec:obs_density_results}
In Figures \ref{barro_1} and \ref{barro_e}, we  show the inferred stellar mass surface densities for each snapshot, as well as the intrinsic value taken directly from the simulation. Stellar mass surface densities are calculated within the central $1\,\rm{kpc}$ and $R_e$ ($\Sigma_1$ and $\Sigma_e$, respectively) for a number of observer inclinations. From Figure \ref{barro_1}, we see that at $z=2.76$ and $z=2.25$, the inferred $\Sigma_1$ shows consistency with the empirically-derived relations of \cite{Barro2017} for all four galaxies. This is in line with the reasonable agreement found for the $R_e-M_\star$ relation. The measured $1\,\rm{kpc}$ surface densities are slightly larger than the intrinsic values. This is due to the overestimation of $M/L_\mathrm{B}$ for halos A2, A4, and A8 at this redshift. At both  $z=2.76$ and $z=2.25$, all inferred surface densities remain consistent with one of the empirical relations. By $z=1.25$, the intrinsic and inferred surface densities are too high for their stellar mass, compared to the observational relations, for all but halo A8.\\
\indent In Figure \ref{barro_e} we show the same relation, but with $\Sigma_1$ replaced by $\Sigma_e$. We find similar behavior to the $R_e-M_\star$ relation, as expected given that $\Sigma_e$ depends on the measurement of $R_e$. Galaxies A4 and A8 show consistency with the star-forming $\Sigma_e-M_\star$ relation derived by \cite{Barro2017}, and galaxies A1 and A2 lie within $1\sigma$ of the quiescent relation. The consistency becomes worse with decreasing redshift, with $R_e$ staying broadly constant at $\sim1\,\rm{kpc}$ while stellar mass increases. By $z=1.25$, all galaxies apart from A8 are too dense. At $z=1.25$, each halo's $\Sigma_e$ is effectively the same as its $\Sigma_1$, with $\Sigma_e$ differing from the empirical relation by a factor of $\sim10$, except for halo A8 which has an $R_e$ that is closer to the size-mass scaling relation (see Figure \ref{vdW}(d)). We will discuss possible reasons for this in Section \ref{sec:discussion}.

\section{Discussion}\label{sec:discussion}
We have attempted to derive an observer's view of the sizes and stellar mass surface densities of massive, intermediate redshift galaxies simulated using FIRE-2 physics. The haloes we study have masses $M_{\rm{halo}}\sim10^{12.5}\,\rm{M_{\odot}}$ and host central galaxies with stellar masses of $\sim10^{11}\,\rm{M_{\odot}}$ at $z=2$. These simulations include recipes for stellar feedback, implemented within a resolved, multi-phase ISM. This is unlike many simulations that match observed central densities via implementations of AGN feedback alongside a much simplified, sub-grid ISM model. The unprecedented resolution of the FIRE-2 simulations enables us to probe the limits of stellar feedback in the extreme environments of the inner regions of massive galaxies.\\
\indent We find that the sizes and surface densities of these simulated massive galaxies are generally within a factor of two of the intrinsic values, calculated directly from the simulations. Across all haloes and redshifts, the median offset between the inferred effective radius and the intrinsic half-mass radius, taken directly from the simulation data, is $0.17\,\rm{dex}$, with inferred radii generally being slightly larger. The standard deviation of the offsets is $0.20\,\rm{dex}$. Both values are consistent with the results of \cite{price2017testing}, who perform similar analysis on FIRE galaxies, but without the detailed radiative transfer modelling that we perform, and find a systematic offset of $\sim0.1\,\rm{dex}$ and a scatter of $\sim0.2\,\rm{dex}$. Across all haloes, redshifts and inclinations, the median offset between $\Sigma_{1,\rm{observed}}$ and $\Sigma_{1,\rm{intrinsic}}$ is $0.04\,\rm{dex}$, with a standard deviation of $0.37\,\rm{dex}$. For  $\Sigma_{e}$, the median offset is $-0.17\,\rm{dex}$, and the standard deviation is $0.52\,\rm{dex}$. While the median offsets are small, the scatter in the offsets is more substantial. This is a concern when considered along with the tightness of empirical relations such as the size-mass relation. We therefore stress the importance of forward-modelling simulations into observational space, for studies that make comparisons between simulated and observationally-inferred scatter in scaling relations. 
\subsection{Comparison to observational relations}
\indent {Having forward-modelled the simulations into the observational plane, we make comparisons with the observationally-derived size-mass relation from \cite{van_der_Wel_CANDELS_size_mass} and the stellar mass-surface density relations from \cite{Barro2017}.} The key result of this paper is that these massive galaxies are, in general, both too small and too dense compared to these empirical relations, with discrepancies increasing towards low redshift. While this is consistent with the study of massive FIRE galaxies performed by \cite{Wellons2019}, less massive FIRE galaxies appear to have more realistic sizes \citep{Garrison2018,Wellons2019}. This could suggest that some piece of physics that is important for massive galaxies is missing from our simulations;
possibilities of such additions will be discussed later in this section. 

\subsection{Uncertainties in observational techniques}
\indent Before discussing possible improvements to the FIRE model, one important point is that the observational relations that we compare to are themselves uncertain. Inferring galaxy effective radii from observations can be difficult: both intrinsic uncertainties about the mass-to-light radio and its constancy or radial dependence across the galaxy, and observational limitations such as the smearing effects of the PSF, limit the robustness of conclusions. Recently, \cite{suess2019half}, argued that color gradients bias the inference of half-mass radii from half-light radii, driving the bulk of the apparent evolution of the size-mass relation. These color gradients are dependent on a number of galaxy properties, including galaxy mass, size, surface density and color, and are not trivial to account for in observational studies. \cite{suess2019half} propose that spatially-resolved SED modelling (e.g. dividing the galaxy into concentric annuli, which are fitted individually) can enable more robust inference of half-mass sizes from multi-band imaging. This approach was adopted by \cite{Mosleh2017} in their study of the evolution of the sizes of star-forming and quiescent galaxies from $z=2$ to $z=0$. \\
\indent We have attempted to circumvent these observational uncertainties by casting our simulated galaxies into `observer space' and making the same assumptions. Nevertheless, our method could be extended to derive stellar mass and effective radii in a more sophisticated manner. \cite{price2017testing} estimate half-mass radii following the approach of \cite{Szomoru2013}, which is also tested by \cite{suess2019half}. The \cite{Szomoru2013} approach uses rest-frame u-band and g-band imaging to constrain possible mass-to-light-ratio gradients and construct color-based stellar mass profiles. This approach yields half-mass radii that are, on average, $\sim25\%$ smaller than rest-frame g-band half-light radii. While this detailed analysis is particularly important for galaxies with strong color gradients, we show in Appendix \ref{sec:appendix_comp} that we are able to recover the intrinsic half-mass radii exceptionally well by simply using effective radii (perhaps because the FIRE-2 galaxies analyzed here are broadly disk-like, with shallow colour gradients), and therefore adopt a simpler strategy.\\
\indent {Rather than infer total stellar mass, we opt to use the intrinsic stellar masses calculated directly from the FIRE-2 simulations.} \cite{price2017testing} derive this quantity by fitting SPS models to the mock photometry using the FAST code \citep{Kriek2009}. They show that stellar masses are recovered extremely well over a wide stellar mass range ($10^{9.5}<M_{*}/\rm{M_{\odot}}<10^{11.25}$), with a median offset of $\log_{10}(M_{*,\rm{recovered}}/M_{*,\rm{intrinsic}})=-0.06\,\rm{dex}$ and a scatter of order $0.1\,\rm{dex}$ over all projections. Thus, introducing stellar mass fitting into our methodology would likely only increase the scatter in our relations very slightly, and we opt to maintain simplicity in this work. \\ 
\indent In this work, we have adopted simple techniques used in the majority of observational studies. Therefore, our results should be similarly susceptible to the biases that affect real observations; in short, if our simulation was well-matched to the galaxies in the real Universe, we would expect our results to be wrong in the same way, and therefore match observations. Therefore, the lack of agreement between our synthetic observations and empirical relations strongly implies that there is some physics missing from the simulation. In the following subsection, we will speculate on where our simulation might be falling short of reality.

\subsection{Possible physical causes of overcompactness}
While the massive simulated galaxies presented in this paper appear to be more compact than observed galaxies of similar stellar mass at the same redshift, one important point to note is that less massive FIRE-2 galaxies do not suffer the same overcompactness \citep{El_Badry_2016}. One likely reason for the overcompactness of the massive FIRE-2 galaxies is the lack of AGN feedback in our simulations. We know from observations that AGN exert feedback on their host galaxies. It is seen directly via radio jets, observable in their strong radio synchotron emission, and via X-ray bubbles and cavities (see the review by \citealt{Fabian2012}). Recent years have also seen increasing amounts of direct observational evidence of `quasar mode' feedback, including observations of high velocity galactic outflows that cannot be attributed to starburst events (see e.g. \citealt{AGN_feedback_Sturm_2011, MRK231_feedback_Rupke_2011,Cicone2014,Fiore2017}). These outflows and their observational signatures have been modelled analytically and in idealized simulations \citep[e.g.][]{Faucher2012,Zubovas2012,Costa2014,Nims2015,Richings2018,Richings2018b}. Motivated by this, and by the need to explain a number of empirical results including the sharp break in the stellar mass function at high masses and the quenching of massive galaxies, many galaxy formation simulations now include some form of AGN feedback \citep[e.g.][]{Springel2005methods,Springel2005red_ellipticals,Dubois2014,Dubois2016,Hirschmann2014,Vogelsberger2014,Schaye2015,Weinberger2017,Weinberger2018,Dave2019}.\\
\indent A number of recent studies have shown that AGN feedback has an impact on galaxy sizes. As discussed in the introduction, \cite{choi2018} perform two sets of cosmological hydrodynamical simulations, one without black holes or AGN feedback (no-AGN runs), and one with AGN feedback in the form of winds and X-ray radiation. The galaxies simulated with AGN have larger half-mass radii at fixed stellar mass. In their simulations, AGN feedback quenches star formation, transforming compact blue galaxies into compact red ones. These quiescent galaxies have lower gas content than their star-forming counterparts in the no-AGN simulations, and subsequently undergo gas-poor mergers that lead to extended stellar envelopes. In addition, fast AGN-driven winds can `puff up' the central region of a galaxy. Differences between the sizes of galaxies in the two simulations become apparent around $z=2$, when in-situ star formation becomes quenched. From around this time, galaxies with AGN evolve more steeply in the mass-size plane than those without AGN. By $z=1$, around half of the galaxies with AGN have become quenched, while those without AGN remain star-forming. The quenched galaxies are clearly separated from the star-forming galaxies in the size-mass plane. Similar results are found by \cite{Dubois2016}, who study lower-stellar-mass galaxies. They show that galaxies simulated with AGN (both heating and jet mode feedback) display larger sizes than their no-AGN counterparts above $M_{*}\sim10^{10}\,\rm{M_{\odot}}$, with the differences increasing with stellar mass for both star-forming and quiescent galaxies, and order-of-magnitude differences by $z=0$. \\
\indent The galaxies with AGN feedback simulated by \cite{choi2018} also show lower $\Sigma_{1}$ values, with an offset of $\sim0.3\,\rm{dex}$ from the no-AGN runs below $z=1$, driven by gas and stellar mass-loss. This is due to high gas accumulation within the central region, with subsequent formation of dense stellar cores. Note, however, that their simulated quenched galaxies do still lie above observationally-derived stellar mass-surface density relations. \cite{Dubois2016} show consistent results. No-AGN galaxies display cuspy centers, whereas massive galaxies with AGN are cored, with flatter central stellar mass densities and a less significant in-situ stellar mass component. \\
\indent These results suggest that the lack of AGN feedback within the FIRE-2 simulations could be one reason for the compact sizes and overdense cores of our galaxies. Our galaxies occupy similar parameter space in the size-mass and density-mass plane to those simulated by \cite{choi2018} without AGN feedback at $z\sim1$ (their galaxies are well-matched to ours, also having $M_{*}\sim10^{11}\,\rm{M_{\odot}}$ at $z=2$). According to their results, our massive FIRE-2 galaxies should be quenched by around $z=1$, rather than continuing to form stars as they do in our simulations. Encouragingly, \cite{angles2017BH_fire} showed that black holes transition to a rapid growth phase when the central stellar potential deepens and star formation becomes less bursty (see also \citealt{Bower2017}, Byrne et al. in prep.). This happens roughly at the time that galaxies exceed $M_{*} \sim \rm{a\,\,few\,\,times\,\,} 10^{10}\,\rm{M_{\odot}}$ and may correspond to the virialization of the inner CGM \citep{Stern2020}. At this stage, an additional source of feedback is required to regulate central densities. Future simulations should address the detailed balance between the higher central densities required for efficient black hole growth and the role of black hole feedback in suppressing central densities.\\

\section{Conclusions}\label{sec:conclusions}
In this paper we have explored the sizes and surface densities of simulated massive galaxies drawn from the FIRE-2 zoom-in simulations \citep{angles2017BH_fire}, which include black hole accretion but not AGN feedback. These simulations model various stellar feedback processes directly within a multi-phase ISM. Thus, the sizes and surface densities of the simulated galaxies can be used to test the efficacy of the feedback model. We focus on the redshift range $1<z<3$, where stellar mass in the Universe is assembling most rapidly. We have modelled the observable sizes of four massive ($M_{*}\sim10^{11}\,\rm{M_{\odot}}$ at $z\sim2$) galaxies, using radiative transfer techniques to include the reddening effects of a realistic dust distribution. We then convolved our images with typical filter profiles and an {\it{HST}}-like PSF, to create mock observations. From these mock observations, we attempted to derive physical properties, mirroring the attempts of observational studies. We base our {{stellar mass surface density}} measurements on well-established observational techniques, which convert an observed color (in our case, $B-R$) to a mass-to-light ratio \citep{bell2001_m_l_relation}. Sizes are derived using a popular S\'ersic profile fitting package, which can successfully reconstruct surface brightness profiles. Our estimates of galaxy sizes and surface densities are generally within a factor of two of the intrinsic quantities, which are inferred directly from the simulations. \\
\indent With the goal of understanding the limitations of our AGN-free simulation, we have compared the inferred sizes of massive FIRE-2 galaxies to the empirical scaling relations derived by \cite{van_der_Wel_CANDELS_size_mass} and \cite{Barro2017}. While the simulated massive galaxies are relatively consistent with empirical size-mass \citep{van_der_Wel_CANDELS_size_mass} and surface density-mass scaling relations \citep{Barro2017} at $z\gtrsim 2$, they significantly diverge from both relations by $z=1.25$. Below $z=2$, the simulated galaxies are too compact compared to observed galaxies at the same redshift, by up to a factor of $10$. The simulated galaxies also become too dense towards low redshifts, with mass surface densities lying well above empirical relations. The most extreme offsets are seen for $\Sigma_{e}$ (rather than $\Sigma_{1}$), due to the added effects of the very small derived $R_{e}$ values. Neither of these offsets can be attributed to purely observational effects, such as sky orientation. \\
\indent The under-predicted sizes and stellar mass surface densities at $z<2$, combined with the fact that less massive FIRE-2 galaxies have been shown to reproduce observationally expected sizes \citep{El_Badry_2016}, indicate that there is some physics missing from these simulated massive galaxies. AGN feedback is expected to play a role in the sizes and star-formation rates of massive galaxies like these, and could be responsible for the discrepancies with observations. We will explore this possibility further with a new suite of simulations that include AGN feedback (Wellons et al., in prep.).

\section*{Data Availability}
The data underlying this article will be shared on reasonable request to the corresponding author. Additional data including simulation snapshots, initial conditions, and derived data products are available at https://fire.northwestern.edu/data/.

\section*{Acknowledgements}
{ We thank the anonymous reviewer for detailed suggestions that helped improve the content and clarity of the paper.\\} 
\indent This work was initiated as a project for the Kavli Summer Program in Astrophysics held at the Center for Computational Astrophysics of the Flatiron Institute in 2018. The program was co-funded by the Kavli Foundation and the Simons Foundation. We thank them for their generous support.  TP would like
to thank all of the participants of the summer program
including Marius Ramsoy, Ulrich Steinwandel and Daisy
Leung for stimulating discussion. \\
\indent TP acknowledges funding from the Future Investigators in NASA
Earth and Space Science and Technology (FINESST) Fellowship, NASA grant 80NSSC19K1610. RKC acknowledges funding from the John Harvard Distinguished Science Fellowship and thanks Sandro Tacchella for helpful discussions. The Flatiron Institute is supported by the Simons Foundation. DAA was supported in part by NSF grant AST-2009687. RF acknowledges financial support from the Swiss National Science Foundation (grant number 157591). CAFG was supported by the NSF through grants AST-1517491, AST-1715216 and CAREER award AST-1652522, by NASA through grant 17-ATP17-0067, by STScI through grant HST-AR-14562.001, and by a Cottrell Scholar Award from the Research Corporation for Science Advancement. SW was supported by the CIERA Postdoctoral Fellowship Program (Center for Interdisciplinary Exploration and Research in Astrophysics, Northwestern University) and by an NSF Astronomy and Astrophysics Postdoctoral Fellowship under award AST-2001905.\\
\indent The simulations were run using XSEDE (TG-AST160048, TG-AST140023), supported by NSF grant ACI-1053575, and Northwestern University's compute cluster `Quest', as well as using NASA HEC allocations SMD-16-7561 and SMD-17-1204. Support for PFH was provided by NSF Collaborative Research Grants 1715847 \&\ 1911233, NSF CAREER grant 1455342, NASA grants 80NSSC18K0562, JPL 1589742. Numerical calculations were run on the Caltech compute cluster ``Wheeler,'' allocations FTA-Hopkins supported by the NSF and TACC, and NASA HEC SMD-16-7592. The data used in this work were, in part, hosted on facilities supported by the Scientific Computing Core at the Flatiron Institute.
This research has made use of the SVO Filter Profile Service, supported from the Spanish MINECO through grant AyA2014-55216.

\bibliography{references}

\appendix\label{sec:appendix}
\section{Mass-to-light ratio versus (B-R) color}\label{sec:appendix_ml}
In Section \ref{sec:sizes_surface_densities}, we describe the process of inferring stellar mass surface densities from our mock observations. In order to do this, we re-derived the \cite{bell2001_m_l_relation} relation (to update it to the AB magnitude system) using the \fsps code \citep{conroy2009_fsps, conroy2010_fsps}, with a \cite{Kroupa2001} IMF. We computed $M/L_\mathrm{B}$ for a modelled stellar population as a function of its $B-R$ color, for various e-folding timescales. Following \cite{bell2001_m_l_relation}, we acquired the $M/L_\mathrm{B}$ and $B-R$ colors of each \fsps track for each e-folding timescale. The $M/L_\mathrm{B}$ and $B-R$ colors of interest correspond to the period of time in which the synthetic stellar population is $12\,\rm{Gyr}$ in age. We fitted these points and derive the relationship:
\begin{equation}
\log(M/L_\mathrm{B})=1.04(B-R) -0.29 \label{bell_fit}
\end{equation}
This new relationship is similar to that derived by \cite{bell2001_m_l_relation}, with a change in the overall normalisation due to the updated SSP models and the use of the AB magnitude system. We use this relation to infer mass-to-light ratios for our FIRE-2 galaxies from the $B-R$ colors output by {\skirt}. We are able to recover intrinsic mass-to-light ratios (using the stellar mass direct from the simulations) very well using this method (typically to within a factor of two).


\section{Comparison of intrinsic and derived sizes and surface densities}\label{sec:appendix_comp}

\begin{figure}[h!]
\includegraphics[scale=0.53]{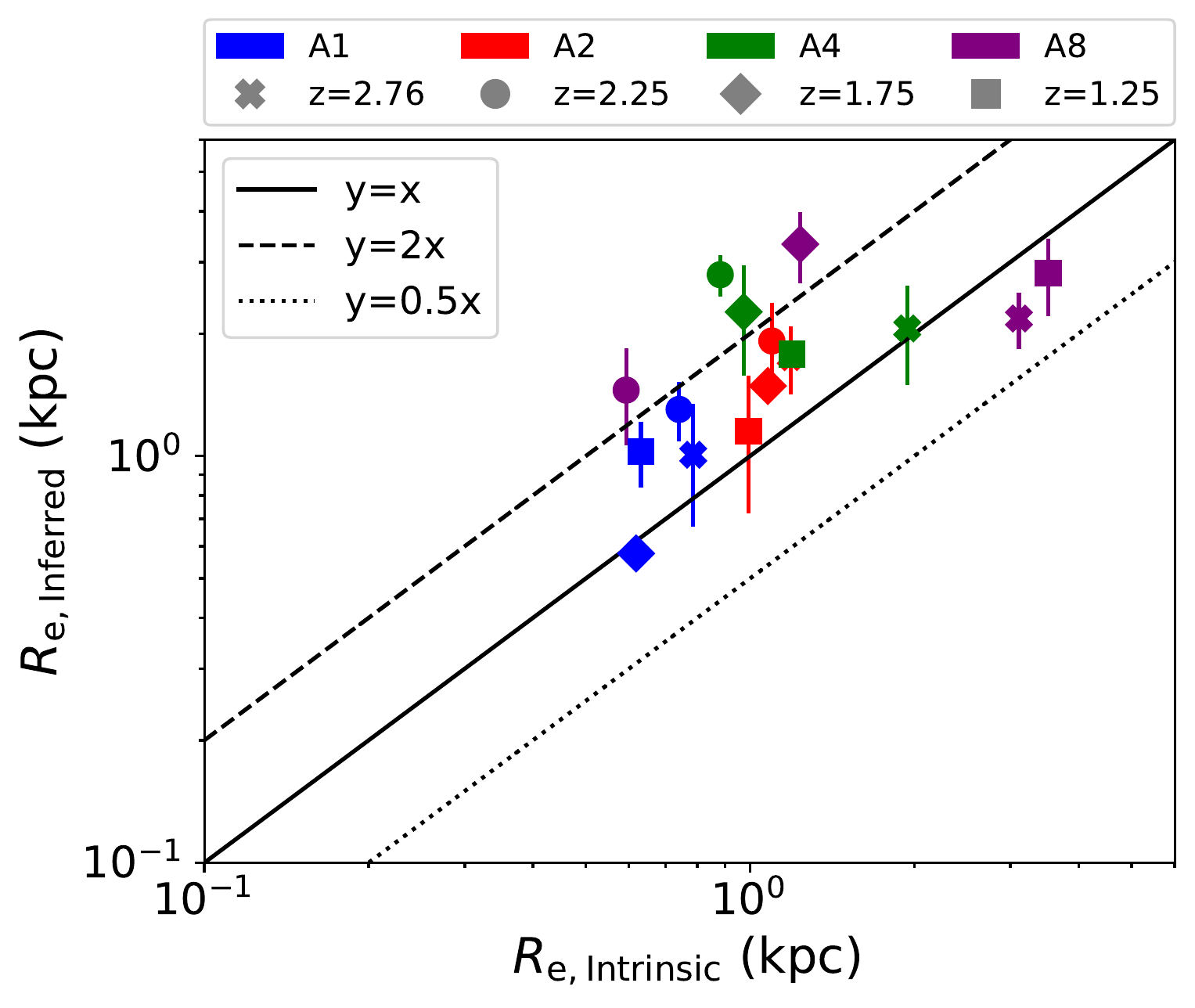}
\caption{The effective radii inferred from S\'ersic fits to our synthetic images, against the intrinsic half-mass radii, measured directly from the simulations. The error bars on our observed sizes are the $1\sigma$ uncertainties calculated using five different sky orientations. The solid black line shows the 1-1 relation, and the dashed/dotted lines show a factor of two offset from this relation. Inferred effective radii tend to be slightly larger than the intrinsic half-mass radii, but the majority of our estimates recover the intrinsic size to within a factor of two.}
\label{size_vs_size}
\end{figure}

\begin{figure}[]
\includegraphics[scale=0.53]{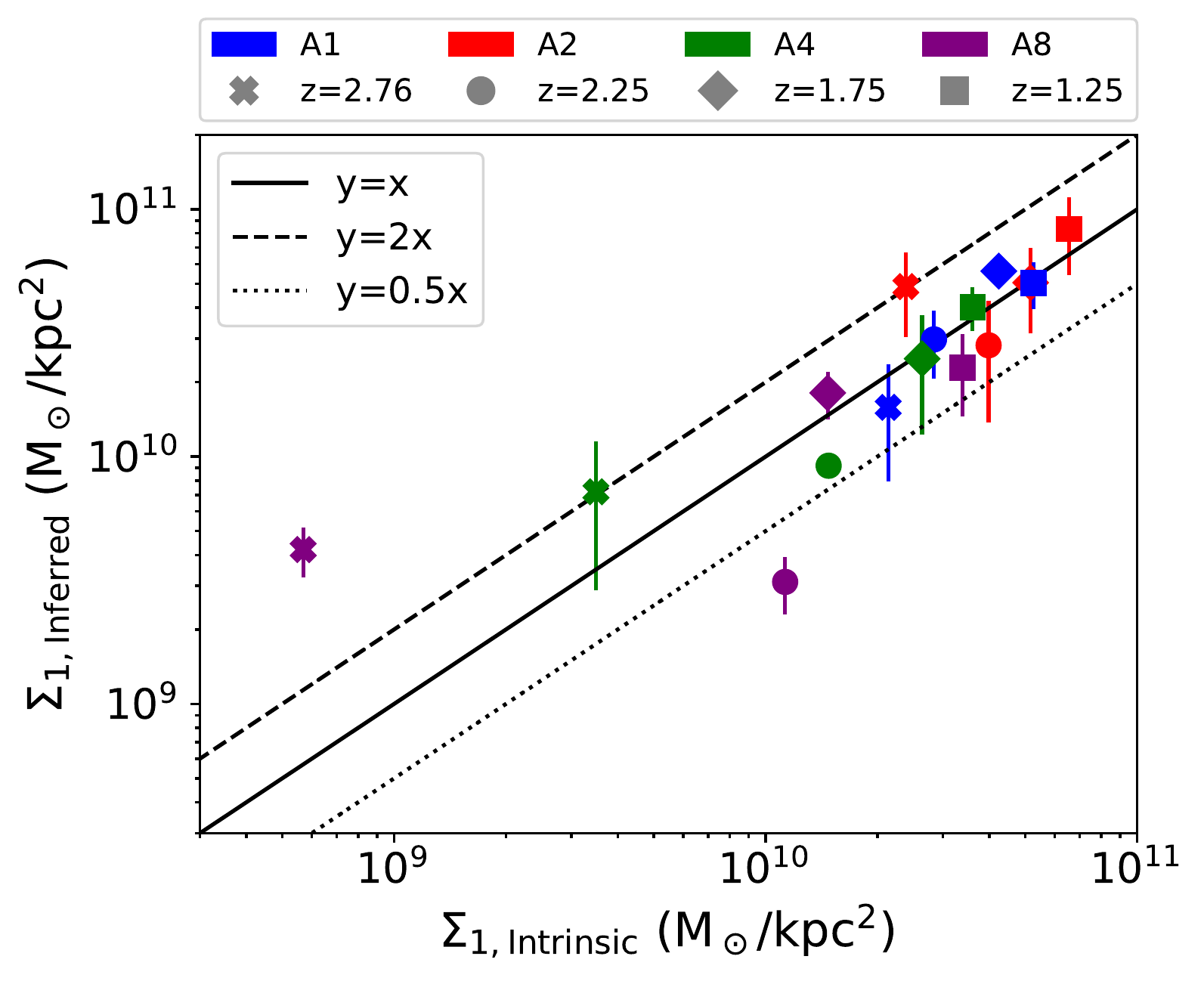}
\includegraphics[scale=0.53]{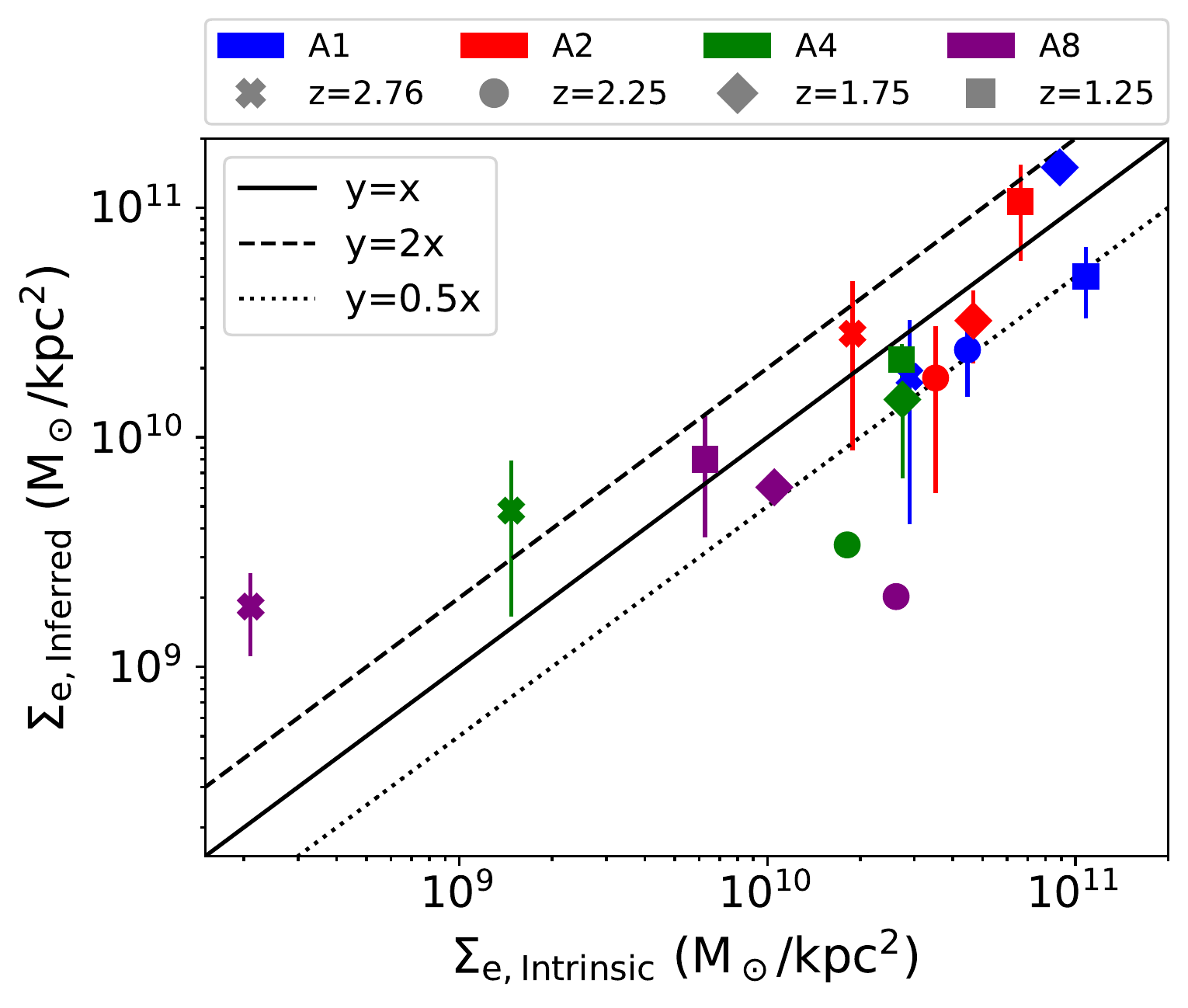}
\caption{The stellar mass surface density within the central $1\,\rm{kpc}$ (upper panel) and $R_{e}$ (lower panel), inferred from our synthetic images, against the same quantities measured directly from the simulations. The error bars on our inferred surface densities are the $1\sigma$ uncertainties calculated using five different sky orientations. The solid black line shows the 1-1 relation, and the dashed/dotted lines show a factor of two offset from this relation. The inferred stellar mass surface densities are, on average, lower than the intrinsic values, but the two values tend to agree within a factor of two. {This overall agreement is in part due to the offsets in M/L ratio and radius coincidentally canceling one another out.}}
\label{density_vs_density_1kpc}
\end{figure}

\bsp	
\label{lastpage}
\end{document}